\documentclass[11pt]{article}

\usepackage[margin=1in]{geometry}
\usepackage{amsmath,amssymb,amsthm}
\usepackage{microtype}
\usepackage{enumitem}
\usepackage{graphicx}
\usepackage{float}
\usepackage{authblk}
\usepackage[hidelinks]{hyperref}

\title{Ripple: An Open, AI-Formalized Lean~4 Framework for Computing with CRNs}
\author[1]{Ho-Lin Chen\thanks{Supported in part by NSTC (Taiwan) grant
113-2221-E-002-204-MY3.}}
\author[2]{Xiang Huang\thanks{Supported in part by Department of Energy EXPRESS
grant DE-SC0024278.}}
\affil[1]{Department of Electrical Engineering, National Taiwan University,
Taiwan\\ \texttt{holinchen@ntu.edu.tw} \quad
\href{https://orcid.org/0000-0002-6171-9962}{\textsc{orcid}\,0000-0002-6171-9962}}
\affil[2]{Department of Computer Science, University of Illinois Springfield,
USA\\ \texttt{xhuan5@uis.edu} \quad
\href{https://orcid.org/0000-0002-4815-6130}{\textsc{orcid}\,0000-0002-4815-6130}}
\date{}

\begin{document}
\maketitle

\begin{abstract}
We present \textbf{Ripple}, an open, AI-formalized Lean~4 framework for the
mathematics of computing with chemical reaction networks (CRNs) --- one
extensible, machine-checked development that gathers several strands of the
field into a single setting, and is built to grow. It formalizes: the theory
of \emph{which real numbers a CRN can compute} --- a single Lean definition of
real-time CRN computation, the class of reals it captures, and the compilation
pipeline (a GPAC / polynomial-ODE layer, a dual-rail compiler, and four stages
down to large-population protocols) that realizes them, built so that adding a
new number is a plug-in; three landmark population-protocol majority algorithms
--- approximate, exact, and self-stabilizing exact majority; the
stochastic-to-deterministic bridge, through three machine-checked versions of
Kurtz's mean-field theorem; and two classical Turing-completeness results ---
Bournez--Gra\c{c}a--Pouly for the deterministic GPAC and
Soloveichik--Cook--Winfree--Bruck for stochastic CRNs. Each of these is
verified to depend on exactly the three Mathlib foundational axioms, with no
\texttt{sorry}. Along the way the formalization repaired genuine, previously
unnoticed gaps in published proofs --- a compilation step that can transiently
leave the unit interval, and an algebraic-number construction that silently
rests on Catalan's conjecture --- and surfaced a sharp open problem about when
a holonomic series admits an exact, rational-data polynomial-ODE encoding. The whole
development is open and every proof is kernel-checked, so the results can be
independently re-verified; and because it was written predominantly by AI
agents using only publicly available models, the workflow that produced it can
be reproduced with the same public toolchain.
\end{abstract}


\section{Introduction}
\label{sec:intro}

Interactive theorem provers have lately moved from a curiosity at the edge of
mathematics to a working tool inside it. Lean~4 and its library Mathlib now
carry machine-checked proofs once thought out of reach --- the Liquid Tensor
Experiment \cite{oScholze22} and the polynomial Freiman--Ruzsa conjecture
\cite{jGGMT25} --- and AI agents have begun proving theorems autonomously,
settling open Erd\H{o}s and OEIS problems \cite{oTsooth26} and disproving the
Erd\H{o}s unit-distance conjecture \cite{oOpen26}. Those breakthroughs used
proprietary internal models; the work here was done entirely with
\emph{publicly available} ones (Claude Opus~4.6--4.8, Claude Fable~5, GPT-5.4--5.6), so the
AI-assisted workflow is reproducible by anyone.

We bring this to molecular programming. \textbf{Ripple}
(\url{https://github.com/zinan-huang/Ripple}) is an \emph{open, AI-formalized}
Lean~4 framework for the mathematics of computing with chemical reaction
networks, covering the theory of CRN-computable real numbers, the
population-protocol majority algorithms, the Turing-completeness of both the
deterministic (GPAC/BGP) and stochastic (SCWB) models, and the supporting
mathematical infrastructure --- continuous-time Markov chains, Kurtz's
mean-field theorem, Frobenius regular-singular theory, and concentration
inequalities. These are exact, layered, and spread across many papers --- the
kind of mathematics where formalization earns its place by catching gaps that
informal proofs leave behind.

\paragraph{The formalized theory.} As an initial thread Ripple formalizes the
theory of \emph{CRN-computable real numbers}, building on three prior papers
\cite{cHuKlLa19,cHuaHul22,cCheHua26} and covering the full ladder of models:
\begin{itemize}[leftmargin=1.4em]
  \item \textbf{Continuous / deterministic.} The GPAC / polynomial-IVP form,
  CRN mass-action dynamics, the CRN-computable reals, dual-rail compilation,
  and the large-population-protocol (LPP) refinement, under one shared Lean
  definition of ``CRN computes $\alpha$''.
  \item \textbf{Stochastic / discrete.} The Angluin--Aspnes--Eisenstat 3-state
  \emph{approximate majority} protocol with its full $O(n\log n)$ convergence
  proof \cite{jAnAsEi08a}, the deterministic correctness chain of the Doty et
  al.\ optimal \emph{exact majority} protocol \cite{cDEGSUS21}, and the silent
  self-stabilizing exact majority protocol of Kanaya et al.\ \cite{cKESOI25}.
  \item \textbf{Turing completeness.} The Bournez--Gra\c{c}a--Pouly (BGP)
  construction \cite{jBoGrPo17} showing that polynomial IVPs can simulate
  arbitrary Turing machines ($\approx 170$ Lean files), and the
  Soloveichik--Cook--Winfree--Bruck stochastic CRN universality theorem
  \cite{jSCWB08} --- the encoding of Turing machines into
  stochastic CRNs ($\approx 90$ Lean files).
  \item \textbf{Kurtz's theorem.} The mean-field limit theorem linking the two
  regimes, making precise how the deterministic ODE is the $N\to\infty$
  limit of the stochastic CRN \cite{jKurt70}.
\end{itemize}

\paragraph{Four things formalization buys.} Formalizing this theory does four
things informal mathematics cannot.
\begin{enumerate}[leftmargin=1.6em]
  \item \textbf{It is reliable.} A machine-checked proof leaves no step for a
  reader to fill in: every inference is checked by Lean's small trusted kernel,
  so trust reduces to that kernel plus three foundational axioms. We verified,
  via Lean's axiom-dependency check, that the core constructions ($\pi/4$,
  $\zeta(3)$, Catalan, LPP, Kurtz) depend on exactly those and nothing more ---
  no \texttt{sorry}, no hidden hypotheses (\S\ref{sec:trust}).
  \item \textbf{It exposes gaps.} The formalization uncovered genuine, fixable
  gaps in published constructions --- in the LPP main theorem and in an
  urn-model argument among them --- each caught as a
  Lean compile error or explicit counterexample (\S\ref{sec:gaps}).
  \item \textbf{It is open and extensible.} Adding a computable number is a
  plug-in; more broadly the foundation --- the GPAC core, the CTMC/Kurtz layer,
  the compilation pipelines --- is a base others can build new threads on
  (\S\ref{sec:reuse}).
  \item \textbf{It proves new things --- and finds new problems.} Ripple settles
  the CRN-computability of $\zeta(3)$ unconditionally by the Fermi--Dirac
  integral route, and machine-checks the Frobenius and holonomic
  generating-function analysis of Ap\'ery's operator --- which, however, does
  not yield a PIVP computing $\zeta(3)$ exactly from rational data
  (\S\ref{sec:new}); the same series recipe
  applied to the modular $1/\pi$ series of Ramanujan surfaces a sharp open
  problem, its series anchor being a fixed point of the drive (\S\ref{sec:pi}).
\end{enumerate}

\paragraph{Organization.} \S\ref{sec:reuse} is an overview of \emph{what the
Lean development contains} --- a map of the framework and the threads it
carries. \S\ref{sec:models} then develops the model ladder in full, from the
GPAC / CRN continuum, through the foundational constants and the LPP compilation
pipeline, to the CTMC layer, Kurtz's theorem, and the LPP$\to$Kurtz bridge from
finite-horizon convergence to readout correctness. \S\ref{sec:gaps} records the gaps the formalization exposed;
\S\ref{sec:new} and \S\ref{sec:pi} then give the new $\zeta(3)$ and $\pi$
constructions in detail, with both the informal arguments and the shape of the
Lean proofs. Finally \S\ref{sec:method} reports how the formalization was
actually done, and \S\ref{sec:trust} delimits the trust footprint.

\section{An Open Framework: Foundations, Threads, and Extension}
\label{sec:reuse}

Ripple is built to be \emph{open}: a reusable foundation that others can
extend, or borrow from to build their own developments. It is best read not as
a closed result but as an \emph{initiative} --- an attempt that carries
multiple threads of research forward, on infrastructure deliberately
made general. Two layers make this concrete. Figure~\ref{fig:repo} is a map of
the repository --- each pillar with its size and what it formalizes --- showing
both the reusable foundation and the threads (completed and in progress) built
on it.

\begin{figure}[!htbp]
\footnotesize
\setlength{\tabcolsep}{3pt}
\renewcommand{\arraystretch}{1.1}
\begin{tabular}{@{}l l p{0.42\linewidth}@{}}
\hline
\textbf{\texttt{Ripple/} pillar} & \textbf{files\,/\,lines\,/\,defs} & \textbf{what it formalizes} \\
\hline
\texttt{Core/} & 15 / 7.2k / 152 & GPAC/PIVP, bounded-time complexity, CRN pipeline \\
\texttt{ODE/} & 1 / 0.3k / 5 & Lyapunov / scalar convergence barriers \\
\texttt{DualRail/} & 18 / 14k / 358 & dual-rail encoding of polynomial dynamics \\
\texttt{LPP/} & 43 / 95k / 2.1k & large-population protocols \cite{cHuaHul22}; algebraic construction; QBee pipeline; compilation stages \\
\texttt{Number/} & 21 / 14k / 565 & $e,\pi,\ln 2,\gamma$, Catalan $G$, Dottie, $\zeta(3)$ (Fermi/Ap\'ery) \\
\quad\texttt{Number/Frobenius/} & 23 / 57k / 2.1k & Frobenius regular-singular theory; Ap\'ery conifold \\
\quad\texttt{Number/Modular/} & 30 / 27k / 1.3k & modular forms, $\Phi_{41}$, CM-163, $j(\tau_{163})$ \\
\quad\texttt{Number/Hypergeom./} & 3 / 5k / 173 & Clausen, Picard--Fuchs Wronskian, Chowla--Selberg; Ramanujan reduction \\
\texttt{CTMC/} & 11 / 35k / 1.2k & DTMC/CTMC, density process, Doob bound, absorbing states \\
\texttt{Kurtz/} & 9 / 3.6k / 78 & Kurtz's mean-field theorem (three versions); log-horizon tube \\
\texttt{Probability/} & 3 / 1.1k / 28 & Bennett exponential-moment lemma; discrete Freedman inequality \\
\texttt{sCRNUniv./} & 92 / 45k / 3.4k & stochastic CRN Turing completeness \cite{jSCWB08} \\
\texttt{BoundedUniv./} & 214 / 161k / 3.6k & GPAC Turing completeness \cite{jBoGrPo17}; bounded extensions (in preparation) \\
\texttt{Analysis/} & 1 / 0.3k / 5 & stable Gr\"onwall lemma \\
\hline
\multicolumn{3}{@{}l}{\emph{population-protocol formalizations (majority)}} \\
\texttt{PopulationProtocol/} & 435 / 303k / 8.3k & \emph{exact majority} (Doty et al.); \emph{approximate majority} \cite{jAnAsEi08a}; \emph{self-stabilizing} \cite{cKESOI25} \\
\hline
\textbf{Total} & \textbf{$\approx$920 / $\approx$768k / $\approx$23.4k} & \\
\hline
\end{tabular}
\caption{Map of the Ripple framework. ``defs'' counts top-level
\texttt{theorem}/\texttt{lemma}/\texttt{def} declarations. The upper block is
the core development; the lower block groups the population-protocol
majority formalizations. Notable additions since submission: \texttt{Probability/}
(Bennett/Freedman), \texttt{sCRNUniv./} (the Soloveichik--Cook--Winfree--Bruck
stochastic CRN universality theorem), \texttt{BoundedUniv./} (the
Bournez--Gra\c{c}a--Pouly GPAC Turing completeness construction, plus
bounded extensions in preparation), and a substantially expanded
\texttt{PopulationProtocol/} pillar now incorporating all three majority
formalizations (Doty et al., Angluin--Aspnes--Eisenstat, and Kanaya et al.).}
\label{fig:repo}
\end{figure}

\paragraph{Foundational infrastructure (reusable, and new to Lean).} A large
part of Ripple is general-purpose and independent of any particular target
number. It includes the GPAC/PIVP core and the single
\texttt{CertifiedBoundedTimeComputable} type (\S\ref{sec:models}) that pins down
what ``computes $\alpha$'' means; the dual-rail compiler; the four-stage LPP
compilation pipeline; and---built from the ground up---a continuous-time Markov
chain layer (\texttt{DTMC}, \texttt{CTMC}, the density process) together with
three machine-checked versions of \emph{Kurtz's mean-field theorem}. These are
foundational definitions and theorems: a verified CTMC theory and a formal
Kurtz limit did not previously exist in Mathlib, and they are reusable far
beyond this paper --- by anyone formalizing stochastic chemical kinetics,
population dynamics, or mean-field limits, whether or not they care about
computable numbers.

\paragraph{Built to extend.} Because the CRN-computable class is a Lean type, adding
a new computable number is a plug-in: supply a PIVP, prove boundedness and an
exponential modulus, invoke the pipeline; a definition proved once is reused
everywhere, the type system guaranteeing the meaning never silently drifts
between theorems written years apart. More broadly, the framework is open in
the strong sense: the foundational layer (the GPAC core, the CTMC/Kurtz theory,
the compilation pipelines) is a base on which others can mount entirely new
threads of research, or which they can take piecemeal for their own.

\section{The Model Ladder}
\label{sec:ladder-overview}
\label{sec:models}

This section develops the model ladder sketched in the introduction into a
self-contained account of what Ripple formalizes and how the pieces
connect: the analog model (\S\ref{sec:ladder-gpac}), CRN-computability
(\S\ref{sec:ladder-firstfloor}), the foundational constants and Catalan
(\S\ref{sec:ladder-constants}), the LPP compilation pipeline
(\S\ref{sec:ladder-lpp}), the stochastic CTMC layer with Kurtz's mean-field
theorem (\S\ref{sec:ladder-kurtz}), and the LPP$\to$Kurtz connection
(\S\ref{sec:ladder-twolimits}).

Throughout: a CRN in its mass-action limit is a polynomial ODE system
\(\dot x_i = p_i(x) - x_i\,d_i(x)\) with \(p_i, d_i\) having non-negative
coefficients; dropping the sign structure gives a PIVP \(\dot y = p(y)\);
and ``\(\alpha\) is CRN-computable'' means a coordinate satisfies
\(|y_k(t) - \alpha| \le C e^{-t}\).

\subsection{The analog model: GPAC, PIVP, and the mass-action CRN}
\label{sec:ladder-gpac}

The continuous model underneath everything is Shannon's \emph{General
Purpose Analog Computer} (GPAC), a circuit of constant, adder, multiplier,
and integrator elements. Shannon proved a GPAC computes exactly the
\emph{differentially algebraic} functions; equivalently, the components of a
solution to a \emph{polynomial initial value problem} (PIVP)
\[
 \dot y = p(y), \qquad y(0) = y_0 \in \mathbb{Q}^d, \qquad
 p \in \mathbb{Q}[y_1,\dots,y_d]^d.
\]
Long believed strictly weaker than Turing computation (the Gamma function
is not differentially algebraic), the GPAC was shown by Bournez,
Campagnolo, Gra\c{c}a, and Hainry (2007) to compute all computable
functions on computable compact intervals, polynomial-time equivalent to
the Turing machine: the polynomial vector field is universal.

A chemical reaction network is the same mathematics with ``concentration''
for ``voltage'': a reaction \(A + B \to C + D\) under mass-action kinetics
fires at a rate proportional to the product of reactant concentrations, so
each species' rate of change is a polynomial. The mass-action CRN is a PIVP
wearing a sign constraint --- production terms are non-negative, and every
negative monomial in \(\dot x_i\) must carry \(x_i\) as a factor (the
Hárs--Tóth ``lack of negative cross-effects'' condition). Hárs and Tóth
(1981) proved this condition is exactly what makes a polynomial ODE
realizable by a mass-action CRN, with an explicit one-reaction-per-monomial
mechanism. For Ripple only the easy direction matters: every bounded PIVP
below splits into production \(p_i\) and degradation \(d_i\) with the right
signs, which is what Lean's \texttt{PolyCRNDecomposition} structure records.

\subsection{CRN-computability}
\label{sec:ladder-firstfloor}

A real number \(\alpha\) is \textbf{CRN-computable} if some PIVP with
rational data has a coordinate \(y_k(t)\) that stays bounded and satisfies
\[
 |y_k(t) - \alpha| \le C\, e^{-t} \qquad \text{for all } t \ge 0.
\]
This is the single notion of computability used throughout: a bounded
polynomial system whose output coordinate converges to the target with a
certified exponential modulus.

The source papers use ``bounded CRN computes \(\alpha\)'' in subtly
different informal senses; in Lean there is exactly one carrier,
\[
 \texttt{CertifiedBoundedTimeComputable}\ (d : \mathbb{N})\ (\alpha : \mathbb{R})
\]
(\texttt{Ripple/Core/BoundedTime.lean}), bundling a PIVP of dimension
\(d\), a designated output coordinate, a boundedness witness, and an
explicit exponential-convergence modulus. Every theorem in the project ---
constant constructions, dual-rail compiler, LPP pipeline --- consumes or
produces this one structure, so ``computes \(\alpha\)'' is type-checked to
be identical across results written years apart. A forgetful coercion drops
the certificate when only the weaker existence statement
(\texttt{BoundedTimeComputable}) is wanted.

\subsection{Foundational constants and Catalan's constant $G$}
\label{sec:ladder-constants}
\label{sec:ladder-catalan}

Ripple re-formalizes the elementary constants --- \(e\), \(\pi\) (through
\(\pi/4 = \arctan 1\)), \(\ln 2\), \(\tfrac12 e^{-1}\), the
Euler--Mascheroni \(\gamma\) (through \(\gamma = 1 - \Gamma'(2)\)), and the
Dottie number --- each as a bounded PIVP with an explicit exponential
modulus, all axiom-free \cite{cHuKlLa19}; the only reciprocal any needs is
the bounded state \(1/(1+t)\) (satisfying \(r' = -r^2\)), which keeps them
inside the polynomial-PIVP class. The worked example of the
integral-to-PIVP recipe is Catalan's constant
\(G = \sum_{n\ge0}(-1)^n/(2n+1)^2 \approx 0.9160\), with the Laplace-type
representation \(G = \int_0^{\infty} \tfrac{s\, e^{-s}}{1 + e^{-2s}}\, ds\).
Its only non-polynomial factor is \(1/(1+e^{-2s})\); carrying it as a state
\(W := 1/(1+e^{-2t})\), which satisfies a quadratic ODE
(\cite[Cor.~19]{cHuaHul22}), turns the whole integral into the four-state
polynomial PIVP
\[
 \begin{cases}
 E' = -E, \\
 R' = E - R, \\
 W' = 2\, E^2\, W^2, \\
 G' = R \cdot W,
 \end{cases}
 \qquad
 (E, R, W, G)(0) = \bigl(1,\ 0,\ \tfrac12,\ 0\bigr),
\]
with closed forms \(E = e^{-t}\), \(R = t\, e^{-t}\),
\(W = 1/(1+e^{-2t})\), so \(G(t)\to G\) with modulus
\(|G(t)-G|\le (t+1)e^{-t}\). This is the theorem
\texttt{catalan\_is\_lpp\_computable}
(\texttt{Ripple/Number/CatalanCertified.lean}), verified clean on the same
generic bounded-PIVP-plus-modulus machinery every target uses.

\subsection{The LPP compilation pipeline}
\label{sec:ladder-lpp}

The \emph{large-population protocol} (LPP) model is the CRN model with
bounded populations: instead of arbitrary real concentrations, one tracks
fractions \(x_i = n_i / N\) of a fixed population of size \(N\). The central
result is \texttt{bounded\_crn\_is\_lpp\_computable\_unconditional}
(\texttt{Ripple/LPP/BoundedLPP.lean}), verified clean: \emph{every} bounded
certified PIVP compiles into an LPP computing the same number. It is
strictly stronger than \cite{cHuaHul22}, which implicitly assumed the
certified computation stays inside \([0,1]\) at all times --- a hypothesis
the formalization showed can fail, since the output may transiently
overshoot while converging from above (see \S\ref{sec:gaps}(A)). The repair
is a saturating low-pass filter \(y' = (x - y)(U - y)\) with
\(U \in (\alpha,1)\cap\mathbb{Q}\), which clamps the output into a
user-chosen interval without changing the limit.

The compiler is a four-stage pipeline, each stage a separately verified
transformation:
\begin{description}[leftmargin=1.6em]
 \item[(i) Quadraticization.] Reduce an arbitrary-degree polynomial vector
 field to degree \(\le 2\) via auxiliary product variables
 \(z = x_i x_j\); preserves the solution and modulus, adds coordinates.
 \item[(ii) Bound-to-small-$\lambda$ closure.] Rescale time/rate so all
 reaction rates fit under the LPP budget (``small-\(\lambda\)''), using the
 boundedness certificate to control the rescaling.
 \item[(iii) Triple-product discretization.] Realize the continuous
 quadratic dynamics by population-level interactions, matching mass-action
 products to the bimolecular rates of a population of size \(N\).
 \item[(iv) Pure-LPP embedding.] Embed the discretized system into the
 formal LPP model, producing the \texttt{IsLPPComputable} witness.
\end{description}
A corollary: every algebraic \(\alpha \in [0,1]\) is LPP-computable, via
\(\dot x_1 = \varepsilon\,q(x_1)\) for a rational polynomial \(q\) with
\(\alpha\) as a stable simple root (e.g.\ \(\pm\) its minimal polynomial),
reduced to degree \(\le 2\) by the quadraticization stage. The formalization also repaired a gap
in the Bournez--Fraigniaud--Koegler conservation trick (another instance of
\S\ref{sec:gaps}): the balancing variable
\(\dot x_\delta = -\sum_j \dot x_j\) is not CRN-implementable as written,
and Ripple's balancing-dilation repair recovers it.

\subsection{The CTMC layer and Kurtz's mean-field theorem}
\label{sec:ladder-kurtz}

A real CRN with \(N\) molecules is a stochastic CTMC; the deterministic ODE
is the \(N\to\infty\) limit (Kurtz, 1970). Ripple's \texttt{CTMC/} and
\texttt{Kurtz/} layers formalize this \texttt{sorry}-free. The density
process \(\bar X^N(t) = X(t)/N\) admits the martingale decomposition
\[
 \bar X^N(t)
 = \bar X^N(0) + \int_0^t F\bigl(\bar X^N(s)\bigr)\, ds + M^N(t),
 \qquad
 \mathbb{E}\!\bigl[\sup_{s \le t} \lVert M^N(s) \rVert^2\bigr]
 \le C t / N.
\]
Three \texttt{sorry}-free versions of the limit are proved: weak
(convergence in probability, \(\sup_{t\le T}\lVert\bar X^N-x\rVert
\to 0\)), strong (a.s.\ rate \(O(\log N/\sqrt N)\) via Azuma--Hoeffding +
Borel--Cantelli), and CLT-scale (\(\mathbb E[\sup\lVert\bar X^N-x
\rVert^2]\le C/N\)).

\subsection{The LPP$\to$Kurtz connection: from finite-horizon convergence to readout correctness}
\label{sec:ladder-twolimits}

An LPP computes \(\alpha\) through \textbf{two limits in a definite order}:
\[
 \underbrace{\text{stochastic PP at size } N}_{\text{finite, random}}
 \xrightarrow[\;N \to \infty\;]{\text{Kurtz}}
 \underbrace{\text{mean-field ODE } x(t)}_{\text{deterministic}}
 \xrightarrow[\;t \to \infty\;]{}
 \underbrace{\alpha}_{\text{the target}}.
\]
Kurtz gives the first arrow on \emph{compact} intervals; it says nothing as
\(T\to\infty\). Passing to a genuine readout guarantee requires two
additional hypotheses: (1)~exponential stability of the limit ODE (the
CRN-computability certificate), and (2)~uniform-in-\(N\) late-time
concentration (a Foster--Lyapunov bound forbidding late escape). Ripple
establishes (1) and the compact-interval Kurtz step \texttt{sorry}-free;
hypothesis (2), the uniform-in-\(N\) late-time concentration, it does
\emph{not} discharge. The order-of-limits content is instead encoded
structurally: \texttt{Computes} is defined on the \emph{post}-population-limit
ODE, so hypothesis (2) is bracketed as an explicit modeling assumption rather
than silently asserted, and the claim factors into these links rather than a
monolithic \(\lim_t\lim_N = \lim_N\lim_t\).

A companion result (\texttt{Ripple/Kurtz/MeanFieldLogHorizon.lean}) takes a
first quantitative step past this barrier: composing a stable-Gr\"onwall
shadowing bound, a Freedman martingale tail, and a lifted-stability
certificate, it proves a tube estimate on the \emph{growing} horizon
\([0,\log N/(2\eta)]\) ---
\(\mathbb P\bigl(\sup_{t\le \log N/(2\eta)}\lVert\bar X^N(t)-\psi(t)
\rVert_\infty > C_K\log N/\sqrt N\bigr)\le N^{-p}\) --- with a
Lipschitz-readout corollary bounding \(|R(\bar X^N(T_N))-\nu|\) at the
endpoint. It is stated modulo those three named inputs rather than fully
unconditionally, but it is precisely the finite-\(N\) readout guarantee the
compact-interval theorem does not by itself provide.

\paragraph{The upshot.}
The deterministic constructions (\S\ref{sec:ladder-constants}) and the LPP
pipeline (\S\ref{sec:ladder-lpp}) live entirely on the ODE side; the
CTMC/Kurtz layer (\S\ref{sec:ladder-kurtz}) connects that side back to the
\emph{physical} stochastic CRN with finitely many molecules. This
subsection is the seam: the formal statement that the number an LPP computes
--- defined through finite-\(N\) stochastic dynamics --- is exactly the number
the corresponding PIVP converges to, with the order of the population and
time limits controlled. Without it, ``the LPP computes \(\alpha\)'' and
``the PIVP converges to \(\alpha\)'' would be two unrelated assertions; with
it, they are the same assertion, type-checked.

\section{Population-Protocol Majority Formalizations}
\label{sec:protocols}

Alongside the CRN-computability thread, Ripple formalizes three landmark
majority protocols from the population-protocol literature --- approximate,
exact, and self-stabilizing exact majority --- together making up the largest
pillar of the repository ($\approx 435$ files, $\approx 303$k lines;
Figure~\ref{fig:repo}).

\paragraph{Approximate majority (Angluin--Aspnes--Eisenstat, 2008).}
The classic 3-state protocol \cite{jAnAsEi08a}: blanks adopt the opinion they
meet, opposing opinions cancel into a blank, and with high probability the
population reaches consensus in $O(n\log n)$ interactions (and, provided the
initial margin is $\omega(\sqrt n\log n)$, the consensus value is the initial
majority). Ripple formalizes the $O(n\log n)$ high-probability
\emph{convergence-to-consensus} theorem --- the quantitative time bound, not
merely stable correctness --- in $\approx 6{,}650$ lines of Lean, zero
\texttt{sorry}; the top-level bound is on $\mathbb{P}(\lnot\,\text{consensus})$,
so it is the consensus-time guarantee rather than the separate
majority-correctness claim. The
formalization follows the paper's own route for the central region, the
product-form supermartingale of its Lemma~4, with the exponential weights
$\exp(\mp c/16n)$ of the original replaced by the rational weights
$\alpha_{vb} = (16n+7)/16n$ and $\alpha_{xy} = (16n-5)/16n$. The rational
weights ask slightly less per step than the paper's exponential ones, but in
exchange the two resulting algebraic inequalities are established exactly and
unconditionally, for every $n \ge 1$, rather than for sufficiently large $n$;
the price is a looser decay constant, which costs nothing asymptotically. The
four regional geometric-decay bounds are then combined by a union bound into
the single $O(n\log n)$ high-probability theorem.

\paragraph{Exact majority (Doty et al., 2021).}
The time- and space-optimal \emph{exact} majority protocol \cite{cDEGSUS21},
which decides the majority correctly with certainty using $O(\log n)$ states
per agent. Ripple formalizes the deterministic correctness chain --- the
protocol's phase structure, its invariants, and the stable-correctness main
theorem --- along with the $O(\log n)$ state-count bound. At $\approx 184$k
lines across $\approx 300$ files it is the largest single formalization in the
repository, reflecting the intricacy of the protocol's cancellation,
doubling/halving, and clock phases.

\paragraph{Self-stabilizing exact majority (Kanaya et al., 2025).}
The time- and space-optimal \emph{silent self-stabilizing} exact majority
protocol \cite{cKESOI25}, which must reach a correct, unchanging consensus from
an \emph{arbitrary} initial configuration. Ripple formalizes the paper's four
theorems: the impossibility of self-stabilizing majority without knowledge of
$n$ (\texttt{impossibility\_without\_n}), the space lower bound
(\texttt{space\_lower\_bound}), convergence, and the time upper bound. The
protocol composes with the \textsc{Optimal-Silent-SSR} ranking subprotocol of
Burman, Chen, Chen, Doty, Nowak, Severson, and Xu \cite{cBCCDNSX21}, whose
deterministic convergence proof Ripple formalizes in full ($\approx 16$k lines
in \texttt{BurmanConvergenceFinal.lean}): the ranking field construction, the
phase-2 propagation/reset, and the epidemic-timer branch to consensus are all
machine-checked. The top-level theorem for the concrete composed protocol
(\texttt{P\_EM\_solves\_SSEM\_final}) is unconditional --- no external
hypotheses --- for all $n \ge 4$. The development is $\approx 111$k lines
across $\approx 99$ files.

\section{Turing-Completeness Formalizations}
\label{sec:universality}

Ripple formalizes two classical universality results, one on each side of the
model ladder: the deterministic GPAC/polynomial-IVP construction of
Bournez--Gra\c{c}a--Pouly, and the stochastic-CRN construction of
Soloveichik--Cook--Winfree--Bruck. Both show that a molecular-computation model
is Turing-complete, and both are machine-checked with the standard three-axiom
footprint.

\paragraph{Deterministic: polynomial IVPs simulate Turing machines (BGP,
2017).} Bournez, Gra\c{c}a, and Pouly proved that a polynomial initial-value
problem --- equivalently, a GPAC --- can simulate an arbitrary Turing machine,
so the class of polynomial ODEs is Turing-complete \cite{jBoGrPo17}. Ripple
formalizes the BGP construction in $\approx 173$ Lean files under
\texttt{BoundedUniversality/BGP/}: the smooth encoding of a machine
configuration into real coordinates, the rational periodic clock, the
single-step transition map, and their composition into one polynomial vector
field whose trajectory tracks the machine's computation. The headline theorem
\texttt{bounded\_pivp\_turing\_complete} states that a single rational bounded
PIVP eventually-threshold-simulates an undecidable machine; its
\texttt{\#print axioms} footprint is exactly $\mathsf{propext}$,
$\mathsf{Classical.choice}$, $\mathsf{Quot.sound}$ and nothing more (``clean-3''),
with no \texttt{sorry} and no custom axiom. The bounded-analog-complexity
extensions built on this construction are in preparation.

\paragraph{Stochastic: Turing machines encode into stochastic CRNs (SCWB,
2008).} Soloveichik, Cook, Winfree, and Bruck showed that finite stochastic
chemical reaction networks are Turing-universal: any Turing machine can be
encoded into a stochastic CRN that simulates it with arbitrarily small error
probability \cite{jSCWB08}. Ripple formalizes this in $\approx 92$ Lean files
under \texttt{sCRNUniversality/}, covering three layers: (i) the deterministic
Turing-machine-to-CRN encoding, in which a four-phase bimolecular refinement
network reproduces each machine step as a reachability fact
(\texttt{statePairTransfer\_crn\_reaches\_of\_ctm\_steps}); (ii) the mass-action
stochastic bridge, which constructs the path-space probability measure from the
mass-action step kernels via Mathlib's Ionescu--Tulcea theorem and bounds the
per-step error probability; and (iii) the decidability of Petri-net
coverability, obtained through the backward coverability algorithm with
termination proved from a well-quasi-order argument (Dickson's lemma). The
paper-facing theorems are machine-checked with the same three-axiom footprint.

\section{Gaps Exposed by Formalization}
\label{sec:gaps}

A machine-checked proof has no human reader to fill in the routine steps. Every
logical step must be justified, every case covered, every interaction pattern
analyzed. In the course of formalizing the constructions that make up the Ripple
pipeline --- both new constructions and the published results they build on --- this
discipline has surfaced a number of genuine gaps. In each case the published
\emph{theorem} is correct; the gap lives in a proof sketch, an intermediate
construction, or an unstated hypothesis, and surfaces in Lean as a compile error
or as an explicit counterexample rather than as a footnote.

We document two representative gaps in detail below --- the mathematics of the
original claim, the precise reason it fails or is incomplete, the explicit
counterexample where one exists, and the fix the formalization carries --- and
note that the formalization turned up others beyond these (for instance an
interior-saddle counterexample to a 2024 algebraic-CRN basin-entry step, and an
atomic-initialization subtlety in the exact-majority error-jump paths).

\medskip
\noindent\textbf{The two gaps detailed here.}
\begin{description}[leftmargin=1.7em,itemsep=2pt]
 \item[(A)] \S\ref{sec:gap-lpp} --- the large-population-protocol \cite{cHuaHul22} main
 theorem: the certified CRN output can transiently leave $[0,1]$; saturating
 low-pass filter fix.
 \item[(B)] \S\ref{sec:gap-cassels} --- reaching algebraic numbers of
 \emph{arbitrary degree} in the urn model needs the non-existence of solutions
 to the Catalan equation $x^p-y^q=1$ (Catalan's conjecture); the classical Cassels~(1960) descent
 supplies only \emph{divisibility}, pinning down exactly which case the argument
 must still close.
\end{description}

\subsection{(A) The LPP main theorem can leave $[0,1]$}
\label{sec:gap-lpp}

This gap is in the large-population-protocol (LPP) main theorem \cite{cHuaHul22}. The LPP model is a natural weakening
of the CRN model in which populations are bounded, and the main theorem states
that every bounded certified PIVP can be compiled into an LPP computing the same
real number.

\paragraph{The unstated assumption.}
As originally written, the theorem assumed that the certified CRN computation
stays inside $[0,1]$ \emph{at all times}. This is the natural setting for a
population protocol: a coordinate is read as a fraction of the population, so it
must lie in $[0,1]$. The trouble is that ``the certified CRN converges to
$\alpha \in [0,1]$ with exponential rate'' does not imply ``the output
coordinate never leaves $[0,1]$.'' A coordinate converging to $\alpha$
\emph{from above} can transiently overshoot $1$ (or, symmetrically, dip below
$0$) by a small amount before settling. In paper form this is at most a
footnote; in Lean it is a genuine compile error, because the embedding step
requires the coordinate to be a valid population fraction at \emph{every} time,
not merely in the limit.

\paragraph{The fix --- a saturating low-pass filter.}
The repair is to post-process the (possibly overshooting) output coordinate $x$
through a saturating low-pass filter
\begin{equation}
 y' \;=\; (x - y)\,(U - y),
 \qquad U \in (\alpha, 1)\cap\mathbb{Q}.
 \label{eq:lpp-filter}
\end{equation}
The factor $(x-y)$ is the ordinary low-pass tracking term --- $y$ relaxes toward
$x$ --- and the additional factor $(U-y)$ is a saturation gate: as $y$
approaches the user-chosen ceiling $U < 1$, the gate $(U-y)$ shrinks the drive
to zero, so $y$ cannot cross $U$. The floor is handled separately: the
input $x$ is a mass-action CRN concentration and therefore non-negative at
all times; since $y(0) \ge 0$ and $y' \ge 0$ whenever $y = 0$ and $x \ge 0$,
the filter output stays in $[0, U] \subset [0, 1]$. Since
$U > \alpha$, the equilibrium of the filtered system still has $y \to \alpha$:
the saturation never engages near the limit, so it clamps the output into a
user-chosen interval \emph{without changing the value computed}. The ceiling $U$
is rational, keeping the filtered system inside the rational-coefficient PIVP
category.

\paragraph{The formalized fix.}
The formalized compilation theorem (\S\ref{sec:ladder-lpp})
incorporates the filter \eqref{eq:lpp-filter} and is \emph{strictly stronger}
than the original paper statement: it drops the always-in-$[0,1]$ hypothesis and
proves the compilation unconditionally for every bounded certified PIVP. The
compilation itself is the four-stage pipeline --- quadraticization,
bound-to-small-$\lambda$ closure, triple-product discretization, and pure-LPP
embedding --- and the filter is what makes the final embedding step go through
in all cases. This is the cleanest illustration in the project of formalization
turning an implicit modeling assumption into an explicit, provable, and
generically applicable repair.

\subsection{(B) Arbitrary-degree algebraic numbers in the urn model: a Catalan-conjecture dependency}
\label{sec:gap-cassels}

This last item situates a population-protocol construction against a classical
number-theoretic lemma, and is about being honest about exactly what that lemma
supplies. The connection to population protocols is direct. The \emph{urn model}
(Albenque--Gerin, 2012) is a population protocol in the strict sense: $n$ agents
each carry one of two colours, at each step a fixed number $k$ of them are
sampled and all recoloured by a majority-type rule, and the proportion of one
colour converges, as $n\to\infty$, to a fixed point of a Bernstein polynomial
--- exactly the large-population-protocol notion of \emph{computing} a number in
$[0,1]$ (\S\ref{sec:ladder-lpp}), specialized to a single two-colour species and
$k$-way interactions. This submodel computes a dense set of algebraic numbers
(while excluding almost every rational), and the headline claim is that it
reaches algebraic numbers of \emph{arbitrary degree}. That claim is what carries
the gap: its proof runs, at one step, into a \emph{Catalan-type} Diophantine
question --- it needs the \emph{non-existence} of solutions to the Catalan equation
$x^p - y^q = 1$ in prime exponents $p,q$ (beyond the lone Catalan exception
$3^2-2^3=1$), i.e.\ Catalan's
conjecture (Mih\u{a}ilescu's theorem). To pin down exactly
what is available toward that, we formalized the classical Cassels (1960)
elementary descent for the equation: $45$ kernel-clean declarations, zero
\texttt{sorry}.

The formalization clarified a point easy to blur in prose, and the point is not
small. Cassels' elementary descent does \emph{not} prove non-existence of
solutions; what it yields is a \emph{divisibility} conclusion --- in a putative
solution $x^p - y^q = 1$ with odd prime exponents, each exponent prime
divides the \emph{opposite} base: $p \mid y$ and $q \mid x$.
Promoting that divisibility to outright non-existence is exactly the strength of
\textbf{Catalan's conjecture}, settled only in 2002 by Mih\u{a}ilescu's theorem
\cite{jMiha04}; the elementary descent does not deliver it. By formalizing
the descent we pinned down \emph{precisely} the residual case --- the step from
``the primes divide as stated'' (which Cassels gives) to ``no solution exists''
(which the urn-model argument needs) --- and showed it to be that deep theorem,
not a routine lemma. The urn model's arbitrary-degree result is therefore sound
(Catalan's conjecture is now a theorem), but it rests, at this one step, on a
genuinely hard result that the elementary presentation hides --- a dependency
only the formalization made explicit.

\section{The $\zeta(3)$ Constructions}
\label{sec:zeta3}
\label{sec:new}

This section develops the $\zeta(3)$ constructions in detail. Ap\'ery's
constant
\[
 \zeta(3) \;=\; \sum_{k\ge 1} \frac{1}{k^3}
 \;=\; 1.2020569031595942\ldots
\]
is first certified as CRN-computable via its Fermi--Dirac integral
representation (routine, but unconditional and machine-checked); the rest of
the section then develops the methodologically novel \emph{series-encoding}
route and the formalization-discovered obstruction that prevents it from
delivering an exact rational-data construction.

\paragraph{CRN-computability via the Fermi integral.}
$\zeta(3)$ has the integral representation
\begin{equation}
 \int_0^{\infty} \frac{x^2}{1+e^{x}}\,dx \;=\; \tfrac32\,\zeta(3),
 \label{eq:zeta3-fermi}
\end{equation}
which Ripple encodes by the same integral-to-ODE move used for
$e,\pi,\ln 2,\gamma$ and Catalan's $G$: a five-state bounded polynomial PIVP
with rational initial data (\texttt{Ripple/Number/AperyFermi.lean}). The
theorem \texttt{apery\_fermi\_is\_crn\_computable} is unconditional, axiom-free,
$0$ \texttt{sorry}. Methodologically this adds nothing new; it settles
CRN-computability so the rest of the section can ask the harder question.

\paragraph{The series-encoding question.}
\textbf{Can $\zeta(3)$ be read off the coefficient sequence $a_n$ and its
holonomic generating function $A(z) = \sum a_n z^n$, encoded as a polynomial
ODE?} The recipe --- take a holonomic sequence, encode the generating
function's linear ODE as a polynomial vector field, recover the target from a
connection coefficient at a regular singular point --- is not specific to
$\zeta(3)$; it is the seed of a research program (\S\ref{sec:zeta3-outlook}).
Ap\'ery's accelerated series is the cleanest first place to exhibit it. We
develop the full series-encoding pipeline below, with Frobenius analysis
machine-checked end to end ($\approx 57$k lines, $0$ \texttt{sorry}), and then
report the obstruction: a one-dimensional neutral symmetry that the
formalization exposed (\S\ref{sec:zeta3-neutral}), which shows that rational
truncated seeding yields $\zeta(3)+c$ rather than $\zeta(3)$ exactly --- the
same barrier that fully blocks the $\pi$/Ramanujan construction of
\S\ref{sec:pi}.

\subsection{The Meaning of ``CRN-computable,'' and the $\zeta(3)$ Difficulty}
\label{sec:zeta3-prelim}

Recall (\S\ref{sec:models}) that a \textbf{polynomial initial value problem}
(PIVP) of dimension $d$ is an autonomous system
\[
 \dot{y} = p(y), \qquad y(0) = y_0 \in \mathbb{Q}^d, \qquad
 p \in \mathbb{Q}[y_1,\dots,y_d]^d,
\]
and that this is the mass-action form of a chemical reaction network (CRN)
once the sign structure $\dot{x}_i = p_i(x) - x_i d_i(x)$ is respected. A real
$\alpha$ is \textbf{CRN-computable} if some coordinate $y_k(t)$ of such a system
stays bounded on $[0,\infty)$ and satisfies $|y_k(t) - \alpha| \le C e^{-t}$ ---
the same notion as in \S\ref{sec:models}. The elementary constants
$e,\pi,\ln 2,\gamma$ and the algebraic numbers are all CRN-computable this way.

The constant $\zeta(3)$ is a genuinely interesting target. Unlike $\zeta(2) =
\pi^2/6$ or the even zeta values, it has \emph{no elementary closed form}: it
is not a known rational multiple of $\pi^3$, and its arithmetic nature beyond
irrationality (Ap\'ery 1978) is still open. The series encoding we develop below
starts from Ap\'ery's accelerated \emph{series}, whose generating function is
holonomic, and encodes that generating function's ODE as a polynomial vector
field. The Frobenius analysis is fully machine-checked; the obstruction to
exact rational seeding is reported in \S\ref{sec:zeta3-neutral}.

\subsection{The Ap\'ery Recurrence and Generating Function}
\label{sec:zeta3-apery}

This route formalizes the combinatorial and differential-algebraic spine of
Ap\'ery's irrationality proof. The deepest piece --- the conifold
connection-coefficient analysis, which leans on Frobenius theory of regular
singular points --- is machine-checked; the residual obstacle is a neutral
symmetry in the polynomial encoding, reported in \S\ref{sec:zeta3-neutral}.

\paragraph{The two Ap\'ery sequences.}
The Ap\'ery numbers are defined combinatorially by
\[
 a_n \;=\; \sum_{k=0}^{n} \binom{n}{k}^2\binom{n+k}{k}^2,
 \qquad a_0 = 1,\ a_1 = 5,\ a_2 = 73,\ a_3 = 1445,
\]
and a companion sequence $b_n$ (the ``harmonic companion'') is pinned by
$b_0 = 0$, $b_1 = 6$. Both satisfy the \textbf{three-term recurrence}
\begin{equation}
 (n+1)^3\,u_{n+1} \;=\; (2n+1)(17n^2+17n+5)\,u_n \;-\; n^3\,u_{n-1},
 \qquad n \ge 1.
 \label{eq:apery-rec}
\end{equation}
Ap\'ery's fundamental fact is that
\[
 \frac{b_n}{a_n} \;\longrightarrow\; \zeta(3),
 \qquad
 \Bigl|\frac{b_n}{a_n} - \zeta(3)\Bigr| = O\bigl((\sqrt2-1)^{8n}\bigr)
 = O\bigl((17-12\sqrt2)^{2n}\bigr),
\]
exponentially fast (roughly three correct digits per term), which is precisely
the rate that drives the irrationality criterion.

\paragraph{F1 --- the recurrence via a pointwise Zeilberger witness.}
The first Lean target, \textbf{F1}, is the recurrence \eqref{eq:apery-rec}
itself for the combinatorial $a_n$. The proof is a creative-telescoping
(Zeilberger) certificate: one exhibits a rational \emph{witness} $W(n,k)$ such
that the summand $c(n,k) = \binom{n}{k}^2\binom{n+k}{k}^2$ obeys a pointwise
telescoping identity
\[
 P_2(n)\,c(n+1,k) + P_1(n)\,c(n,k) + P_0(n)\,c(n-1,k)
 \;=\; W(n,k+1) - W(n,k),
\]
where $P_2(n) = (n+1)^3$, $P_1(n) = -(2n+1)(17n^2+17n+5)$, $P_0(n) = n^3$.
Summing over $k$ collapses the right-hand side telescopically to boundary
terms, yielding \eqref{eq:apery-rec}. In Lean the subtlety is the case
analysis on the summation regime: the binomial supports differ between
$c(n-1,k)$, $c(n,k)$, $c(n+1,k)$, so the witness identity must be checked
separately in the three regimes
\[
 k \le n-2, \qquad k = n-1, \qquad k = n,
\]
(plus the trivially-zero tail). Crucially, once the algebraic identities are
assembled, each regime closes by a \emph{single} \texttt{linear\_combination}
call --- the polynomial certificate is exact, so no inequality or limiting
argument is needed --- and the whole of F1 is axiom-free.

\paragraph{F1$'$ --- the harmonic-correction companion recurrence.}
The companion $b_n$ is not given by a single binomial sum but by a
\emph{harmonic-weighted} one. Ripple formalizes this as \textbf{F1$'$} via the
decomposition
\[
 b_n \;=\; H_3(n)\,a_n + d_n,
 \qquad H_3(n) = \sum_{m=1}^{n}\frac1{m^3},
\]
where $H_3$ is the order-$3$ harmonic number and $d_n$ is the correction
term. F1$'$ proves that $d_n$ satisfies its own companion recurrence (the same
left-hand operator as \eqref{eq:apery-rec} with an explicit harmonic forcing),
which is what makes $b_n$ a solution of \eqref{eq:apery-rec} for $n \ge 1$
while $b_0 = 0, b_1 = 6$ encodes the irrationality data. This is the discrete
shadow of the inhomogeneity that appears in F2 below.

\paragraph{F2 --- the formal Ap\'ery ODE for the generating functions.}
Form the ordinary generating functions $A(z) = \sum_{n\ge 0} a_n z^n$ and
$B(z) = \sum_{n\ge 0} b_n z^n$. Because the coefficients are $P$-recursive,
the generating functions are holonomic (D-finite): they satisfy a linear ODE
with polynomial coefficients, obtained from \eqref{eq:apery-rec} by the
\textbf{Euler operator} $\theta = z\,d/dz$ (which acts as $\theta z^n = n
z^n$, converting polynomial-in-$n$ recurrences into polynomial-in-$\theta$
operators). The result is the third-order \textbf{Ap\'ery differential
operator}
\begin{equation}
 L[u] \;=\; p(z)\,u''' + q(z)\,u'' + r(z)\,u' + s(z)\,u,
 \label{eq:apery-op}
\end{equation}
with
\[
 \begin{aligned}
 p(z) &= z^2(1-34z+z^2), &\qquad q(z) &= z(3-153z+6z^2), \\
 r(z) &= 1-112z+7z^2, &\qquad s(z) &= -5+z.
 \end{aligned}
\]
\textbf{F2} states the two formal identities
\begin{equation}
 L[A] \;=\; 0, \qquad L[B] \;=\; 6.
 \label{eq:apery-AB}
\end{equation}
The $A$-series solves the \emph{homogeneous} equation; the $B$-series solves
the \emph{inhomogeneous} one with constant term $6$. This asymmetry is not a
bug: $a_n$ and $b_n$ obey the same recurrence \emph{for $n \ge 1$}, but the
initial data $(b_0,b_1) = (0,6)$ violates the $n=0$ instance (which would
require $u_1 = 5u_0$), and the defect $b_1 - 5b_0 = 6$ propagates as a
constant forcing term. The action of $L$ on $B$ telescopes to a boundary term
determined by $b_0,b_1$, yielding exactly the constant $6$. In Lean, F2 is a statement about \emph{formal} power series:
$L[A]$ and $L[B]-6$ are shown to have all coefficients zero, by reducing each
coefficient identity back to the recurrence \eqref{eq:apery-rec} and the F1$'$
companion. No analysis (convergence, radius) enters F2; it is pure
coefficient algebra, and it is \texttt{sorry}-free.

\subsection{The Frobenius Infrastructure (Framework-Level Investment)}
\label{sec:zeta3-frobenius}

What F2 does \emph{not} supply is the analytic passage from the formal
identities \eqref{eq:apery-AB} to the exponential-rate convergence of the
coefficient ratio $b_n/a_n \to \zeta(3)$. That passage requires the Frobenius
theory of regular singular points, which Mathlib currently lacks. We explain
the mathematics, then state what is and is not formalized.

\paragraph{Singular structure of the Ap\'ery operator.}
The leading coefficient $p(z) = z^2(1-34z+z^2)$ vanishes at
\[
 z = 0 \qquad\text{and}\qquad z = 17\pm 12\sqrt2,
\]
so the singular points of \eqref{eq:apery-op} are $0$, the \textbf{conifold
point} $z_1 = 17-12\sqrt2 \approx 0.0294$, its conjugate $17+12\sqrt2$, and
$\infty$. Since $a_n \sim C\,\alpha^n n^{-3/2}$ with $\alpha = 1/z_1 =
17+12\sqrt2$, the radius of convergence of $A$ (and $B$) is $z_1$, and $z_1$ is
the dominant singularity. At a \emph{regular} singular point, Frobenius
theory predicts a basis of solutions of the form $(z_1-z)^\rho\cdot(\text{
analytic})$ or with an added $\log$ factor, where the exponents $\rho$ are
roots of the \textbf{indicial polynomial}.

\paragraph{Indicial analysis at the conifold.}
Substituting $y = (z_1-z)^\rho$ into the leading part of \eqref{eq:apery-op}
near $z_1$ gives the indicial equation
\[
 \rho(2\rho-1)(\rho-1) \;=\; 0,
 \qquad\text{roots } \rho \in \{0,\ \tfrac12,\ 1\},
\]
with the corresponding local solutions
\[
 \begin{aligned}
 \varphi_0(z) &= c_0 + c_1(z_1-z) + \cdots & (\rho=0,\ \text{analytic}),\\
 \varphi_1(z) &= \sqrt{z_1-z}\,\bigl(d_0 + d_1(z_1-z)+\cdots\bigr)
 & (\rho=\tfrac12,\ \text{branch point}),\\
 \varphi_2(z) &= (z_1-z)\,\bigl(e_0 + e_1(z_1-z)+\cdots\bigr)
 & (\rho=1,\ \text{analytic}).
 \end{aligned}
\]
The decisive object is $\varphi_1$, the $\sqrt{z_1-z}$ solution, whose second
derivative diverges, $\varphi_1''(z) \sim -\tfrac14(z_1-z)^{-3/2} \to \infty$,
while $\varphi_0,\varphi_2$ and their derivatives stay bounded.

\paragraph{Connection coefficients: $b_n/a_n \to \zeta(3)$.}
Writing $A = \alpha_0\varphi_0 + \alpha_1\varphi_1 + \alpha_2\varphi_2$ and
$B = \beta_0\varphi_0 + \beta_1\varphi_1 + \beta_2\varphi_2$ in the local
basis, the error function $E(z) = B - \zeta A = \sum(b_n - \zeta a_n)z^n$ has
its singular ($\varphi_1$) component
\[
 (\beta_1 - \zeta\,\alpha_1)\,\varphi_1
\]
vanish precisely when $\zeta = \beta_1/\alpha_1$. At that value, $E$ is a
combination of the analytic $\varphi_0,\varphi_2$ only, hence analytic at
$z_1$; its radius of convergence jumps to the next singular point
$z_2 = 17+12\sqrt2 \approx 34 \gg z_1$. The arithmetic content of Ap\'ery's proof is exactly the statement that
this \textbf{connection coefficient} equals the target,
\[
 \frac{\beta_1}{\alpha_1} \;=\; \zeta(3).
\]
Equivalently, since both $A''$ and $B''$ are dominated by the same
$\varphi_1'' \sim (z_1-z)^{-3/2}$ blow-up,
\[
 \frac{B''(z)}{A''(z)} \;=\; \zeta(3) + \frac{E''(z)}{A''(z)}
 \;\xrightarrow{\,z\to z_1\,}\; \zeta(3),
\]
the second-derivative ratio converging at rate $O\bigl((z_1-z)^{3/2}\bigr)$.
This is also the analytic reason the \emph{coefficient} ratio $b_n/a_n$
converges exponentially: the gap between the two radii of convergence forces
$b_n - \zeta(3)a_n = O\bigl((17-12\sqrt2)^{n}\bigr)$.

\paragraph{The formalized content.}
The combinatorial and differential-algebraic layers (F1, F1$'$, F2) are
\texttt{sorry}-free. The passage from the formal ODE \eqref{eq:apery-AB} to the
connection-coefficient identity $\beta_1/\alpha_1 = \zeta(3)$ is carried by the
lemma \texttt{apery\_conifold\_frobenius\_witness}, which is proved (it is not a
\texttt{sorry}) --- but it is \emph{conditional}: it takes the three-halves
conifold bound as a hypothesis and rests on
\texttt{apery\_conifold\_frobenius\_bridge}, a named residual gap. Under those
inputs it yields the exponential modulus $|\rho(t) - \zeta(3)| \le K e^{-\kappa
t}$ for \emph{exact} seeds entering the conifold basin. It does not on its own
certify CRN-computability of $\zeta(3)$: the truncated rational seeds one can
actually use shift the limit to $\zeta(3) + c$ (\S\ref{sec:zeta3-neutral}), so
the unconditional result is supplied only by the integral route
(\texttt{apery\_fermi\_is\_crn\_computable}). This analytic
content is genuinely deep: it is the same Frobenius / monodromy analysis of the
Ap\'ery operator at $\{0, z_1, \infty\}$ that Beukers and others study in
connection with irrationality measures of $\zeta(3)$, which is what makes the
machine-checked version a substantial result rather than a routine lemma. (The
reusable payoff is correspondingly broad: the same conifold-Frobenius machinery
applies to the bulk analog-computability of holonomic generating functions, not
merely to $\zeta(3)$.)

\paragraph{The origin obstruction.}
A complementary feature explains why the bare series route is awkward to run,
and why the construction Lean actually certifies takes a different form. The
point $z=0$ is a \textbf{maximally unipotent monodromy} (MUM) point: all three
indicial exponents at the origin coincide ($\rho = 0,0,0$ for the operator
normalized at $z=0$), producing logarithmic solutions. When the ODE is lifted
to an autonomous polynomial system by the time reparametrization
$dz/d\tau = p(z)$, the origin --- which would carry the rational initial data
--- becomes a \emph{fixed point} of $\dot z = p(z)$, so the flow cannot leave
it. Seeding from a nearby rational $z_0 > 0$ with truncated-Taylor data works
numerically (machine precision; see below), but the arrival time diverges like
$\tau \sim \Theta(1/z_0)$ and the convergence is correspondingly slow. The
$x=\tau^2$ uniformization absorbs the $\sqrt z$ Puiseux branch (indicial roots
become $\{0,0,1\}$) but neither removes the double root / log term nor
unfreezes the origin. This is why the clean, certified construction of
$\zeta(3)$ is the Fermi-integral route
(\texttt{apery\_fermi\_is\_crn\_computable}). The \emph{conifold} route below
does better than the bare series seeding --- it reads $\zeta(3)$ off a ratio of
blow-ups at a non-fixed point, avoiding the frozen origin --- but it still
requires truncated rational seeds, and \S\ref{sec:zeta3-neutral} shows the
truncation shifts the limit to $\zeta(3) + c$.

\subsection{The Derivative-Ratio PIVP and the Periods Connection}
\label{sec:zeta3-derivative}

Although the series route is awkward to seed, the analysis above yields an explicit,
fully polynomial PIVP with \emph{rational} initial data whose readout converges
to $\zeta(3) + c$, the exponentially small neutral-mode offset of
\S\ref{sec:zeta3-neutral} --- a construction worth recording because it is the
concrete object that obstruction lives on, and because it generalizes to the
entire class of periods.

\paragraph{The adaptation law and the unbounded system.}
The mechanism that extracts $\zeta(3)$ is an \emph{adaptation law}: introduce a
state $\zeta(\tau)$ and drive it toward the instantaneous ratio $B''/A''$,
\[
 \dot\zeta \;=\; \varepsilon\,A''\,\bigl(B'' - \zeta\,A''\bigr),
\]
a gradient-style update whose stable equilibrium is $\zeta^\star = B''/A'' \to
\zeta(3)$ (the linearization at equilibrium is $-\varepsilon(A'')^2 < 0$). To
make every right-hand side polynomial we reparametrize time by the drive
$dz/d\tau = p(z)$ --- so $z \to z_1$ exponentially in $\tau$, giving the
adaptation law unbounded time to converge --- and eliminate the third
derivatives using \eqref{eq:apery-AB}, i.e.\ $A''' = (-q A'' - r A' - s A)/p$
and $B''' = (-q B'' - r B' - s B + 6)/p$, the leading $p$ cancelling. The result
is the explicit \textbf{unbounded} eight-state polynomial system
\begin{equation}
 \begin{aligned}
 \dot A &= A'\,p, & \dot A' &= A''\,p, & \dot A'' &= -q A'' - r A' - s A, \\
 \dot B &= B'\,p, & \dot B' &= B''\,p, & \dot B'' &= -q B'' - r B' - s B + 6, \\
 \dot z &= p, & \dot\zeta &= \varepsilon\,A''\,(B'' - \zeta A''), &&
 \end{aligned}
 \label{eq:apery-unbounded}
\end{equation}
with rational initial data $A^{(j)}(z_0), B^{(j)}(z_0)$ (truncated Taylor
evaluations at $z_0 = 1/1000$) and $\zeta(0) = 0$ --- no prior knowledge of the
target. This system converges to within the neutral-mode offset of $\zeta(3)$,
but $A', A'', B', B''$ all diverge as $z \to z_1$, so it is not yet a
\emph{bounded} CRN.

\paragraph{From the derivative ratio to a bounded polynomial system.}
The naive function-value ratio $B(z)/A(z)$ does \emph{not} tend to $\zeta(3)$
(numerically it tends to $\approx 0.35$, contaminated by the analytic
$E(z_1)\neq 0$); the $\zeta(3)$ information lives in the \emph{singular} part,
which is subdominant in $B/A$ but dominant in the derivative ratios. Accordingly
we divide the unbounded system \eqref{eq:apery-unbounded} through by the
fastest-growing variable $A''$, tracking $R = B''/A''$; this gives the bounded
ratio variables
\[
 \alpha = \frac{A'}{A''},\quad
 \sigma_A = \frac{A}{A''},\quad
 \beta = \frac{B'}{A''},\quad
 \rho = \frac{B''}{A''},\quad
 \sigma_B = \frac{B}{A''},\quad
 w = \frac{1}{A''},
\]
all of which stay order-$1$ ($\alpha,\sigma_A,\beta,\sigma_B,w \to 0$, $\rho
\to \zeta(3)$). With the reparametrization $dz/d\tau = p(z)$ and the
abbreviation $Q = q + r\alpha + s\sigma_A$, the quotient rule produces a fully
polynomial system (degree $\le 4$, eight variables including an adaptation
variable $\zeta$):
\[
 \begin{aligned}
 \dot z &= p(z), \\
 \dot\alpha &= p + q\alpha + r\alpha^2 + s\,\sigma_A\,\alpha, \\
 \dot\sigma_A &= \alpha\,p + \sigma_A\,Q, \\
 \dot\beta &= \rho\,p + \beta\,Q, \\
 \dot\rho &= r(\rho\alpha - \beta) + s(\rho\,\sigma_A - \sigma_B) + 6w, \\
 \dot\sigma_B &= \beta\,p + \sigma_B\,Q, \\
 \dot w &= w\,Q, \\
 \dot\zeta &= \varepsilon\,(\rho - \zeta).
 \end{aligned}
\]
The forcing constant $6$ --- the same inhomogeneity from $L[B]=6$ --- enters
the $\dot\rho$ equation explicitly as $+6w$ and is what drives $\rho$ (hence
$\zeta$) toward $\zeta(3)$ rather than a meaningless number. All right-hand
sides are polynomial, all initial conditions (truncated Taylor evaluations of
$A,B$ at a rational $z_0$ such as $1/1000$) are rational, and all variables
are bounded. Integrating from $\zeta_0 = 0$, the adaptation variable matches
$\zeta(3)$ to machine precision ($\approx 4.4\times 10^{-16}$, Radau solver).
As \S\ref{sec:zeta3-neutral} shows, the exact limit is in fact $\zeta(3) + c$
with $c$ far below this precision --- the numerics cannot distinguish the
encoding from an exact one, which is exactly what the formalization detected.

\begin{figure}[H]
\centering
\includegraphics[width=\linewidth]{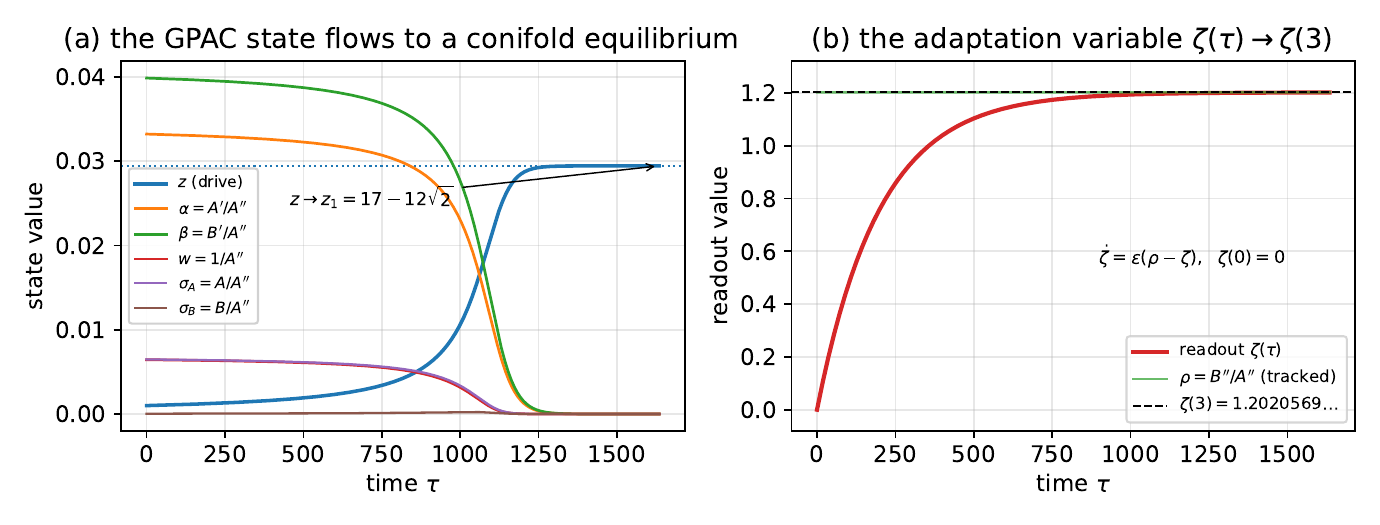}
\caption{The $\zeta(3)$ GPAC as an ODE system --- the bounded eight-variable
polynomial PIVP of \S\ref{sec:zeta3-derivative}, integrated (DOP853) from
rational data at $z_0=10^{-3}$. \textbf{(a)} the projective state flows to a
conifold equilibrium: the drive $z$ relaxes to $z_1=17-12\sqrt2$ while the
bounded coordinates $\alpha=A'/A''$, $\beta=B'/A''$, $w=1/A''$, $\sigma_A=A/A''$
and $\sigma_B=B/A''$ settle. \textbf{(b)} the dedicated \emph{readout}
variable $\zeta(\tau)$, seeded at $\zeta(0)=0$ with no prior knowledge of the
target, climbs to $\zeta(3)=1.2020569\ldots$ under the gradient adaptation law
$\dot\zeta=\varepsilon(\rho-\zeta)$, tracking the conifold ratio
$\rho=B''/A''$. (Behind the scenes $A''$ and $B''$ both blow up like
$(z_1-z)^{-3/2}$; the ratio $\rho$ stays pinned near $\zeta(3)$, which is what
$\zeta$ locks onto.) The machine-checked content of the route
(\texttt{apery\_conifold\_frobenius\_witness}) is that for \emph{exact} seeds
the conifold ratio $\rho$ tends to the connection coefficient
$\beta_1/\alpha_1 = \zeta(3)$; for the truncated rational seeds used here,
\S\ref{sec:zeta3-neutral} shows the true limit is $\zeta(3) + c$, with $c$ below
the numerical precision shown.}
\label{fig:zeta3}
\end{figure}

\subsection{The Lean Development}
\label{sec:zeta3-summary}

It is worth being concrete about what this $\zeta(3)$ work contains, because
the engineering is as much the point as the theorems. The unconditional
certification --- that $\zeta(3)$ is CRN-computable
(\texttt{apery\_fermi\_is\_crn\_computable}) --- comes from the Fermi-integral
route, a routine construction. The heavy investment went into the
\emph{series-route analysis}, which does not itself certify but was not a
transcription of a known proof either: it required building, in Lean~4 and from
scratch, mathematics that Mathlib does not contain --- a Frobenius theory of
regular singular points, the indicial and connection-coefficient analysis at
the Ap\'ery conifold, and the holonomic generating-function machinery for the
Ap\'ery operator. The
\texttt{Ripple/Number/Frobenius/} development alone runs to roughly
\textbf{57{,}000 lines and 2{,}000 declarations} (Figure~\ref{fig:repo}), none
of which previously existed in a proof assistant, on top of the \texttt{Number/}
constant library and the hypergeometric layer. This series-route chain --- the
three-term recurrence (F1), the harmonic companion (F1$'$), the formal Ap\'ery
ODE (F2: $L[A]=0$, $L[B]=6$), and the conifold-Frobenius witness
(\texttt{apery\_conifold\_frobenius\_witness}, conditional on the three-halves
bound and the residual \texttt{bridge}) --- is machine-checked, and is
independent of the Fermi theorem that supplies the unconditional result. But
the series route does not deliver a rational-data PIVP that converges to
$\zeta(3)$ exactly; the next subsection explains why.

\subsection{The Neutral Mode}
\label{sec:zeta3-neutral}

The 8-variable bounded PIVP of \S\ref{sec:zeta3-derivative} is the projectivization
of the linear jet $(A, A', A'', B, B', B'')$ by $A''$: each variable is a
ratio with denominator $A''$ (for instance $\alpha = A'/A''$,
$\rho = B''/A''$, $w = 1/A''$). Because $A$ solves the homogeneous equation
$L[A] = 0$ while $B$ solves the inhomogeneous $L[B] = 6$, the transformation
$B \mapsto B + cA$ (for any constant $c$) preserves the inhomogeneous equation
and hence all eight ODEs. Under it
\[
 \rho \;\mapsto\; \rho + c, \qquad \zeta \;\mapsto\; \zeta + c,
\]
while $\alpha, \sigma_A, z, w$ are unchanged.

The consequence: two rational seeds in the same symmetry orbit produce
trajectories whose limits differ by exactly $c$. The initial data at
$z_0 = 1/1000$ are truncated Taylor evaluations of $A$ and $B$ --- rational,
but not \emph{exact}. The truncation introduces a component along $A$ of size
$c \approx z_0^N$ (exponentially small in the truncation order $N$, but
nonzero), so $\rho(t) \to \zeta(3) + c$, not $\zeta(3)$.

This is not a deficiency of the Frobenius analysis (which is fully proved) but
a structural property of the encoding: $\zeta(3)$ enters the system as a
connection coefficient --- a two-point scattering datum linking the
maximally unipotent monodromy (MUM) point $z = 0$
(\S\ref{sec:zeta3-frobenius}) to the conifold $z_1$ --- and the neutral mode is the degree of
freedom that carries it. At $z = 0$ the true data is rational ($A(0) = 1$,
$B(0) = 0$), which pins $c = 0$; but $z = 0$ is a double zero of the
polynomial drive $p(z) = z^2(1-34z+z^2)$, so no trajectory can depart from it.
This is the \textbf{same MUM barrier} that fully blocks the $\pi$/Ramanujan
construction of \S\ref{sec:pi} --- for both series, the obstruction is the
gap between exact (transcendental) initial values at a regular basepoint and
the rational data the PIVP definition requires.

Numerics at $4.4 \times 10^{-16}$ residual said the construction works;
formalization said the limit is $\zeta(3) + c$. This single discovery is
the clearest illustration in the project of formalization as an instrument
that finds structure informal mathematics overlooks.

\subsection{Outlook: $\zeta(3)$ as the Seed of a Research Program}
\label{sec:zeta3-outlook}

The construction above does more than certify one constant; it fixes a
\emph{recipe} for turning an infinite series into a chemical reaction network,
and it is worth stating the recipe abstractly because that is what generalizes.
Given a sequence $(a_n)$ whose ordinary generating function $A(z) = \sum a_n
z^n$ is \emph{holonomic} (satisfies a linear ODE with polynomial coefficients),
the pipeline is:
\begin{enumerate}[leftmargin=2em]
 \item \textbf{Holonomic ODE.} Obtain the linear ODE
 $\sum_{i} p_i(z)\,A^{(i)}(z) = 0$ from the recurrence for $(a_n)$ (creative
 telescoping / Zeilberger), as in F1--F2 for the Ap\'ery sequences.
 \item \textbf{Polynomial vector field.} Realize $A$ and its derivatives, the
 monomials $z^i$, and the auxiliary quotients as coordinates of a
 \emph{polynomial} initial value problem --- a GPAC, hence (after the dual-rail
 sign discipline) a CRN.
 \item \textbf{Read-off.} Recover the target real number as a limiting value
 or a connection coefficient of the trajectory, and certify the convergence
 rate.
\end{enumerate}
Nothing in steps 1--3 used special features of $\zeta(3)$ beyond holonomicity.
This suggests three concrete directions, which we state as \emph{future work},
not as results:
\begin{itemize}[leftmargin=2em]
 \item \textbf{A general holonomic-series theorem.} Identify hypotheses on a
 holonomic $(a_n)$ under which $\sum a_n$ (or $\lim b_n/a_n$) is provably
 CRN-computable at a stated convergence rate. For $\zeta(3)$ the
 conifold-Frobenius analysis is machine-checked
 (\texttt{apery\_conifold\_frobenius\_witness}), but even there exact rational
 seeding is blocked by the neutral mode (\S\ref{sec:zeta3-neutral}) and the
 CRN-computability conclusion comes from the integral route; the general
 theorem would need both to abstract that conifold-Frobenius argument into
 reusable hypotheses on $(a_n)$ and to resolve the seeding question, ideally
 backed by a general Frobenius theory of regular singular points in Mathlib.
 \item \textbf{Multiple and nested series.} Sums such as multiple zeta values
 $\zeta(s_1,\dots,s_k)$ are governed by holonomic systems in several variables;
 the natural question is whether the same encode-the-ODE move lifts to
 $D$-finite functions of several variables.
 \item \textbf{Multiple integrals.} Period integrals (e.g.\ the Beukers triple
 integral for $\zeta(3)$, or higher-dimensional periods) admit Picard--Fuchs
 ODEs; encoding those Picard--Fuchs ODEs would bring period integrals and
 series under the same generating-function mechanism used here.
\end{itemize}
In this sense $\zeta(3)$ is not the destination but the entry point: it is the
smallest target that cannot be reached by the elementary integral move and
therefore the first that exercises the full series-encoding machinery. The
current status is that the machinery is built and machine-checked on the
$\zeta(3)$ instance; even there the read-off step is blocked by the neutral
mode, so exact CRN-computability rests on the integral route, and the
generalizations above are open.

\paragraph{The recipe stalls here (a lead-in to $\pi$).} The recipe is not
universal, and the next section makes the limitation concrete. Ramanujan's
$1/\pi$ series encodes into an equally elegant polynomial system ---
the holonomic-ODE step goes through beautifully --- yet the construction stalls
for a sharp structural reason: the series' natural anchor is the point $z=0$,
which is at once a singular point of the generating-function ODE and a
\emph{fixed point} of the polynomial drive, so the flow cannot leave it and
exact rational seeding is impossible. $\zeta(3)$ nearly escaped this (up to the
neutral mode of \S\ref{sec:zeta3-neutral}) only because it reads off a
\emph{ratio} of blow-ups at a non-fixed conifold. Whether the
MUM-anchored $1/\pi$ series admits a comparable truncation-free device is, as
far as we know, open --- an inviting problem that \S\ref{sec:pi} lays out, and
that sits squarely on the holonomic-series program above.

\section{The $\pi$ Construction}
\label{sec:pi}

This section is about a single, sharply posed question: can the fast
\emph{modular} $1/\pi$ series of Ramanujan be turned into an exact GPAC/CRN
construction? It is worth separating cleanly what is \emph{classical} from what
this paper does. That the Ramanujan~(1914) series equals $1/\pi$ is a classical
theorem; the molecular-programming community needs no re-proof of it, and Ripple
takes that identity as given (in Lean it is carried as one explicit
complex-multiplication hypothesis, with the surrounding reduction machine-checked
and $0$-\texttt{sorry}). What we contribute is the analog-computation reading: a
clear derivation that encodes the series into a clean six-state polynomial PIVP
(\S\ref{sec:pi-ramanujan}), followed by an account of why the
encoding nonetheless does \emph{not} compute $\pi$ exactly --- its readout series
is anchored at $z=0$, a \emph{singular point} of the holonomic ODE at which
the compiled system cannot be seeded: the reciprocal state blows up there, and
the drives that avoid the blow-up stall there
(\S\ref{sec:pi-obstruction}).

One point should be made and then set aside: that $\pi$ \emph{is}
CRN-computable, exactly and unconditionally, is not in question and is not new.
Like $e$, $\ln 2$, and the Euler--Mascheroni $\gamma$, it is an elementary,
linearly computable target settled in Huang et al.'s earlier work
\cite{cHuKlLa19} (in Ripple, $\pi/4=\arctan 1$ is re-formalized axiom-free as
\texttt{pi\_quarter\_is\_lpp\_computable}). The modular-series question of this
section is a different and harder one --- not whether $\pi$ can be computed at
all, but whether \emph{this particular fast series} admits an exact,
truncation-free encoding.

Throughout, ``computes $\alpha$'' means a polynomial initial-value problem
(PIVP) --- the GPAC normal form $\dot y = p(y)$, $y(0)\in\mathbb{Q}^d$,
$p\in\mathbb{Q}[y]^d$ --- with a designated coordinate $y_k(t)$ such that
$|y_k(t)-\alpha|\le C e^{-t}$, exactly the CRN-computability definition of
\S\ref{sec:models}. The decisive criterion, applied without exception below, is
binary: the continuous-time limit must equal the \emph{exact} target constant. A
family parametrized by a truncation order whose limit is $\pi+\delta$ with
$\delta\neq 0$ does not compute $\pi$, however small $\delta$ may be.

\subsection{Encoding Ramanujan's 1914 series as a six-state GPAC}
\label{sec:pi-ramanujan}

\paragraph{The identity.} In 1914 Ramanujan published a list of seventeen series
for $1/\pi$; the simplest is
\begin{equation}
 \frac{1}{\pi} \;=\; \frac{2\sqrt{2}}{9801}
 \sum_{k=0}^{\infty}
 \frac{(4k)!}{(k!)^4}\cdot\frac{1103 + 26390\,k}{396^{4k}}.
 \label{eq:ramanujan}
\end{equation}
The first term alone already gives $1/\pi$ to eight decimal places, and each
further term adds about another eight. The goal is to read the right-hand side
off the attractor of a polynomial vector field, step by step.

\paragraph{Step 1 --- the coefficients are holonomic.} Write the Ramanujan
coefficient $a_k := (4k)!/(k!)^4$. A direct computation on the ratio of
consecutive coefficients gives the P-recurrence
\begin{equation}
 (k+1)^3\,a_{k+1} \;=\; 4\,(4k+1)(4k+2)(4k+3)\,a_k,
 \label{eq:ram-recurrence}
\end{equation}
so $a_{k+1}/a_k$ is a rational function of $k$ --- the sequence is
\emph{holonomic}, and its generating function $g_\pi(z) := \sum_k a_k z^k$
satisfies a linear ODE with polynomial coefficients.

\paragraph{Step 2 --- the generating function's ODE.} With the Euler operator
$\theta := z\,d/dz$, \eqref{eq:ram-recurrence} translates into the Clausen-type
hypergeometric equation
\begin{equation}
 \theta^3 g_\pi \;=\; 256\,z\,
 \bigl(\theta+\tfrac14\bigr)\bigl(\theta+\tfrac12\bigr)\bigl(\theta+\tfrac34\bigr) g_\pi,
 \label{eq:ram-clausen}
\end{equation}
i.e.\ $g_\pi(z)={}_3F_2(\tfrac14,\tfrac12,\tfrac34;\,1,1;\,256z)$, with singular
points $z=0$, $z=1/256$, $z=\infty$. Converting to ordinary derivatives and
clearing one factor of $z$ gives the form we actually integrate,
\begin{equation}
 z^2(1-256 z)\,g_\pi''' \;=\; 3z(384z-1)\,g_\pi'' + (816z-1)\,g_\pi' + 24\,g_\pi.
 \label{eq:ram-ode}
\end{equation}
At $z=0$ the indicial polynomial is $\rho^3$ --- a triple root, the point of
\emph{maximally unipotent monodromy} (MUM); two coefficients of
\eqref{eq:ram-ode} vanish there, the third at $z=1/256$.

\paragraph{Step 3 --- the readout.} A short manipulation expresses Ramanujan's
series at the evaluation point $z_0:=1/396^4$ as a linear combination of $g_\pi$
and $z g_\pi'$:
\begin{equation}
 \frac{1}{\pi} \;=\; \frac{2\sqrt2}{9801}
 \bigl[\,1103\, g_\pi(z_0) + 26390\, z_0\, g_\pi'(z_0)\,\bigr].
 \label{eq:ram-readout}
\end{equation}
So once the machine can evaluate $g_\pi(z_0)$ and $g_\pi'(z_0)$, \eqref{eq:ram-readout}
hands us $1/\pi$.

\paragraph{Step 4 --- rescale away from the MUM point.} Because $z_0\approx
4.1\times10^{-11}$ sits awkwardly close to the singularity at $0$, set
$w:=z/z_0\in[0,1]$ and $g(w):=g_\pi(z_0 w)$, so $g(1)=g_\pi(z_0)$ and
$g'(1)=z_0 g_\pi'(z_0)$. The chain rule converts \eqref{eq:ram-ode} into an ODE
of the same shape on $[0,1]$,
\begin{equation}
 w^2(1-256 z_0 w)\,g''' \;=\; 3w(384 z_0 w-1)\,g'' + (816 z_0 w-1)\,g' + 24 z_0\, g,
 \label{eq:ram-ode-w}
\end{equation}
the second regular singular point now pushed out to $w=99^4\approx10^8$, far
outside $[0,1]$; the MUM at $w=0$ is unchanged. The readout \eqref{eq:ram-readout}
becomes $\tfrac{1}{\pi}=\tfrac{2\sqrt2}{9801}[1103\,g(1)+26390\,g'(1)]$.

\paragraph{Step 5 --- polynomialize.} Equation \eqref{eq:ram-ode-w} solved for
$g'''$ carries the non-polynomial factor $1/[w^2(1-256 z_0 w)]$, which the GPAC
model forbids. Two standard moves remove it. (i) Carry the reciprocal
$Q := 1/[w^2(1-256 z_0 w)]$ as a new state; by the quotient rule
$Q' = -2w(1-384 z_0 w)Q^2$, which \emph{is} polynomial in $(w,Q)$. (ii) Drive
$w$ from the seed to $1$ by the linear relaxation $\dot w = 1-w$, which multiplies
every $g$-derivative by the polynomial drift $(1-w)$. Finally a closed-loop
inverter $\dot P = 1 - I\,P$, with
$I = \tfrac{2\sqrt2}{9801}(1103\,g + 26390\,g')\to 1/\pi$, converts $1/\pi$ to
$\pi$ with no division.

\paragraph{The six-state system.} Collecting everything, the encoding is the
six-dimensional PIVP with state $X=(w,Q,g,g',g'',P)$:
\begin{equation}
 \begin{cases}
 \dot w \;=\; 1 - w, \\[2pt]
 \dot Q \;=\; -\,2\,w\,(1-384 z_0 w)\,Q^2\,(1-w), \\[2pt]
 \dot g \;=\; g'\,(1-w), \\[2pt]
 \dot g' \;=\; g''\,(1-w), \\[2pt]
 \dot g'' \;=\; Q\,\bigl[\,3w(384 z_0 w-1)\,g'' + (816 z_0 w-1)\,g' + 24 z_0\, g\,\bigr](1-w), \\[2pt]
 \dot P \;=\; 1 - \dfrac{2\sqrt2}{9801}\bigl[\,1103\,g + 26390\,g'\,\bigr]P.
 \end{cases}
 \label{eq:ram-pivp}
\end{equation}
Every right-hand side is polynomial in the six states; the readout is
$\lim_{\tau\to\infty}P(\tau)=\pi$, with $w\to1$ and $I\to1/\pi$ at rate $1$ and
$P$ relaxing at rate $1/\pi\approx0.318$. The result is a polynomial
vector field whose only attractor has $\pi$ in its last coordinate.

\paragraph{Removing the $\sqrt2$.} The lone irrational coefficient is the
$2\sqrt2/9801$ in the inverter. The clean way to remove it is \emph{not} to
carry $\sqrt2$ as an auxiliary state and multiply at the end --- a product of
two approximations compounds their errors --- but to \emph{read off $\pi^2$}
with a rational-coefficient inverter and then take one square root, handling the
inversion and the de-squaring together (the faster-inverter device of
\cite{cCheHua26}). Writing $T = 1103\,g + 26390\,g'$, which is rational in the
states, the readout \eqref{eq:ram-readout} squares to $1/\pi^2 =
(8/9801^2)\,T_\infty^2$, i.e.\ $\pi^2 = 9801^2/(8\,T_\infty^2)$. The squared
inverter
\[
 \dot P_{\mathrm{sq}} \;=\; 9801^2 \;-\; 8\,T^2\,P_{\mathrm{sq}}
 \;\longrightarrow\; \pi^2
\]
has \emph{rational} coefficients throughout and relaxes fast (rate
$8\,T_\infty^2$), and a Newton-style square-root ODE $\dot y = c\,(P_{\mathrm{sq}}
- y^2) \to \pi$ recovers $\pi$. No irrational constant and no error-compounding
product appears anywhere; every coefficient of the system is rational.

\paragraph{Lean status.} Ripple's \texttt{Ramanujan1914.lean}
verifies the recurrence \eqref{eq:ram-recurrence} and the coefficient form of
the ODE \eqref{eq:ram-clausen}. The identity \eqref{eq:ramanujan} rests on a
chain of modular-forms machinery that Ripple formalizes in substantial depth:
the Klein $j$-invariant, the Legendre $\lambda$-function, theta-quartic
identities, the $\Phi_{41}$ modular polynomial with a Sturm certificate
($\approx 3500$ coefficients), and the CM evaluation
$j\!\bigl(\tfrac{1+\sqrt{-163}}{2}\bigr) = -640320^3$
(\texttt{KleinJCM163Statement}, $\approx 27$k lines across 30 files, $0$
\texttt{sorry}). The one residual hypothesis is \texttt{hcm58}, a
Chowla--Selberg normalization step connecting the CM evaluation to the
specific series coefficients of \eqref{eq:ramanujan}. We claim the
\emph{encoding} of \eqref{eq:ramanujan} into the polynomial system
\eqref{eq:ram-pivp} is explicit and verified; the series identity is
conditional only on \texttt{hcm58}.

\subsection{The singular-seed obstruction}
\label{sec:pi-obstruction}

The six-state system \eqref{eq:ram-pivp} falls short of the
standard met by $\zeta(3)$, and the reason is instructive rather than
fatal. The seed of the construction is the value of $g$ and its derivatives near
$w=0$ --- equivalently $z=0$. But $z=0$ is exactly the bad point: it is
\emph{simultaneously}
\begin{itemize}\setlength{\itemsep}{1pt}
 \item a \emph{singular point} of the holonomic ODE \eqref{eq:ram-ode}, the MUM
 point where two leading coefficients vanish; and
 \item a \emph{singular seed} of the compiled system: the reciprocal state
 $Q=1/[w^2(1-256z_0w)]$ of \eqref{eq:ram-pivp} is infinite at $w=0$, so the
 six-state flow cannot be initialized there. Compiling instead through the
 Euler form $\theta=w\,d/dw$ keeps every state finite at $w=0$, but at the
 price of a drive $\dot w \propto w$ for which $w=0$ is a fixed point and
 a trajectory started there never leaves.
\end{itemize}
The construction must therefore start not at the singularity but at a nearby
rational $w_1 = \varepsilon > 0$, seeding $g(w_1), g'(w_1), g''(w_1)$ from a
\emph{truncated} Taylor polynomial of the series. The exact seed values are
transcendental, so rational data forces truncation, and the continuous-time
limit becomes $\pi+\delta$ with $\delta$ rigorously bounded (driven below any
target by taking more Taylor terms) but \emph{not exactly zero}. By the strict
criterion of this paper --- the limit must equal the target on the nose --- the
encoding \eqref{eq:ram-pivp} \emph{approximates} $\pi$ to arbitrary precision but
does not \emph{compute} it.

This is not a defect of the encoding; it is an
\textbf{open problem}: can the modular $1/\pi$ series be realized by a polynomial
PIVP with rational data whose continuous-time limit is \emph{exactly} $\pi$,
without seeding away from the singular point?

The obstruction is the same one the $\zeta(3)$ series route encounters
(\S\ref{sec:zeta3-neutral}): in both cases, the target constant enters the
system as a connection coefficient --- a two-point scattering datum linking the
MUM point to a conifold --- and the polynomial drive has a zero at the MUM
point, making exact rational data dynamically inaccessible. For $\zeta(3)$
the conifold ratio device gets within one neutral direction of the target
($\rho \to \zeta(3) + c$, $c$ exponentially small); for $\pi$ there is no
companion solution and no conifold ratio, so the MUM barrier is complete.
One barrier, two positions on it: $\zeta(3)$ nearly escapes because Ap\'ery's
companion sequence exists; $\pi$'s Ramanujan series has no such companion.
Whether a comparable device exists for MUM-anchored $1/\pi$ series
is, to our knowledge, unresolved.

\section{The Formalization Method}
\label{sec:method}

It is worth describing, plainly, how this body of Lean was actually produced,
because the method is itself part of what we wish to report.

\paragraph{An AI-driven workflow.} The formalization was carried out
predominantly by AI agents. The workflow was: the agents read the source
papers directly; the human authors occasionally supplied a curated reference or
a key mathematical idea; and the bulk of the Lean~4 development --- stating the
definitions, discharging the goals, refactoring, and debugging against the
kernel --- was performed autonomously by the models, iterating against
\texttt{lake build} until each file compiled cleanly.

\paragraph{The division of labor.} The human authors
contributed the \emph{mathematical} constructions --- for instance the
$\zeta(3)$ adaptation law of \S\ref{sec:new}, the bounded-analog-complexity
stratification, and the large-population-protocol compilation pipeline. The
\emph{formalization} of these into machine-checked Lean, and notably the
discovery of the gaps documented in \S\ref{sec:gaps} (which surfaced as Lean
compile errors and explicit counterexamples), was largely the work of the AI
agents. On the Lean side, the CRN-computability / LPP compilation thread drew on
the human authors' own prior constructions and required specific mathematical
input at key steps; the remaining threads --- the population-protocol
formalizations, the CTMC/Kurtz layer, the Frobenius/modular infrastructure,
and the universality formalizations --- were driven almost entirely by the AI
agents, with human contribution limited to direction and review.

\paragraph{Publicly available models --- the methodological point.} All of this
was done with \emph{publicly available} large language models (Anthropic's
Claude Opus~4.6--4.8 and Fable~5, and OpenAI's GPT-5.4--5.6), \emph{not} the proprietary
internal systems behind recent AI-mathematics milestones such as the
autonomous Lean proof search that settled open Erd\H{o}s and OEIS problems
\cite{oTsooth26} or the AI disproof of the Erd\H{o}s unit-distance conjecture \cite{oOpen26}.
The significance is reproducibility: the workflow described here can be run by
anyone with access to these public models. Formalizing a substantial body of
analog-computability theory --- the GPAC/CRN ladder, the LPP pipeline, the
CTMC/Kurtz bridge, and the number constructions --- is now within reach of a
standard, publicly available toolchain, not only of closed in-house systems.

\paragraph{Setup, scale, and how much was human.} For reproducibility we record
the concrete configuration.
\emph{Hardware:} development and orchestration ran on a desktop workstation (a
$24$\,GB Mac mini, Apple M4, 10 cores); the Lean builds, which are
memory-intensive, ran on a shared Linux server (128 cores,
$\approx 503$\,GB RAM, cgroup quota $256$\,GB per user).
\emph{Agents:} the work was driven by two publicly available agentic coding
tools --- \textbf{Claude Code} (Anthropic, running Claude Opus~4.6--4.8 and
Fable~5) and \textbf{Codex} (OpenAI, running GPT-5.4--5.6) --- which read the papers, write
and edit Lean, run \texttt{lake build}, and iterate on the errors.
\emph{Duration:} the formalization has been under active development for
roughly four months (late March to mid-July 2026).
\emph{Scale:} the repository is about $768$k lines of Lean~4 across
$\approx 920$ files; the largest pillars are the population-protocol majority
formalizations ($\approx 303$k lines), the bounded-universality / BGP
construction ($\approx 161$k), the Number pillar including Frobenius and
modular machinery ($\approx 103$k), the LPP pipeline ($\approx 95$k),
and the stochastic CRN universality ($\approx 45$k).
\emph{Human intervention:} on the \emph{mathematical} side, the
CRN-computability / LPP compilation thread drew on the human authors' own
prior constructions and required specific mathematical input at key steps.
The remaining threads --- the population-protocol majority formalizations
($\approx 303$k lines), the bounded-universality / BGP and stochastic CRN
universality constructions ($\approx 206$k lines combined), the CTMC/Kurtz
layer, and the Frobenius/modular infrastructure --- required \emph{no human
mathematical input}: the AI agents read the source papers directly and
produced the formalizations autonomously, with the human contribution limited
to direction and review. On the \emph{Lean} side, the humans wrote
essentially none of the $\approx 768$k lines by hand.

\section{Trust Footprint}
\label{sec:trust}

\paragraph{The axiom audit.}
Ripple has \textbf{no \texttt{axiom}} declarations and \textbf{no
\texttt{sorry}}. Across the headline results --- $\pi/4$, $\zeta(3)$,
Catalan's $G$, the LPP compilation theorem, the three Kurtz mean-field
theorems, the BGP and SCWB universality theorems --- we verified with
Lean's axiom-dependency check (\texttt{\#print axioms}) that each depends on
\emph{exactly} the three Mathlib foundational axioms $\mathsf{propext}$,
$\mathsf{Classical.choice}$, $\mathsf{Quot.sound}$ and nothing more. Two
deeper statements are explicitly conditional, each on a single named
hypothesis carried as a \texttt{Prop} argument: the $\zeta(3)$ conifold
convergence-rate theorem (modulo the neutral-mode bound of
\S\ref{sec:zeta3-neutral}), and the Ramanujan $1/\pi$ identity (modulo one
CM-normalization, \S\ref{sec:pi}).

\paragraph{Acceptance criteria and LLM escape prevention.}
Because the formalization was produced predominantly by AI agents, a
recurring concern is \emph{escape}: the model finds a formally valid proof
that is mathematically vacuous --- an \texttt{axiom} that makes the goal
trivial, a \texttt{sorry} hidden behind an import chain, a
\texttt{native\_decide} that trusts the compiler on a statement the kernel
cannot check, a spurious hypothesis smuggled into a theorem statement (the
model adds a \texttt{Prop} argument that quietly assumes the conclusion,
making the proof trivially true --- a pattern especially common with
GPT-5.4-era agents), or a definition that is technically inhabited but
degenerately so (e.g.\ \texttt{def~A~:=~B} when $A$ was supposed to be an
independent construction of $B$). The formalization playbook enforces five gates against
these:
\begin{enumerate}[leftmargin=1.8em]
  \item \textbf{The \texttt{\#print axioms} gate.} Every deliverable theorem
  is audited: \texttt{\#print~axioms~T} must list exactly the three Mathlib
  foundations. Any appearance of \texttt{sorryAx} or a custom axiom name
  fails the gate. This catches \texttt{sorry} and \texttt{axiom} escapes
  anywhere in the transitive import closure, not just in the file being
  reviewed --- a \texttt{grep sorry} misses sorry-free files that import
  sorry-carrying dependencies. Textual grep is used only as a coarse
  upper bound (it false-positives on docstrings containing the word
  ``sorry''); the authoritative signal is always \texttt{\#print axioms}.
  \item \textbf{The cold-build gate.} A full \texttt{lake build} from a
  clean state (no cached \texttt{.olean} files) is the only reliable
  compilation test. Warm builds reuse stale objects and can mask broken
  imports, namespace collisions, dangling opens after refactoring, and
  cross-file definition drift. Single-file \texttt{lake env lean} passes
  are necessary but not sufficient: a file can compile alone while its
  downstream consumers break. Every merge into the main branch requires a
  cold build to pass.
  \item \textbf{The non-vacuity check.} A clean-3 theorem can still be
  vacuous if its hypotheses are unsatisfiable or its conclusion is
  trivially true. The playbook requires a concrete \emph{witness}: for a
  computability theorem, a numerical integration that matches the certified
  limit; for a universality theorem, a specific machine instance. Spec
  files for AI agents explicitly list \emph{banned identifiers} to prevent
  the shortest-path escape of defining the target in terms of itself.
  \item \textbf{The version-proliferation ban.} AI agents frequently
  produce versioned copies of the same file (\texttt{V3}, \texttt{V4},
  \texttt{Final}, \texttt{Capstone}) rather than fixing the original. Each
  version may carry its own \texttt{sorry}s, polluting the import closure;
  a ``clean'' capstone that transitively imports a sorry-carrying predecessor
  is not clean. The playbook enforces: one theorem, one live file; when a
  new version lands, the old one is deleted or quarantined in the same
  commit.
  \item \textbf{The counterexample-before-attack rule.} Before assigning
  an agent to prove a residual goal, the playbook requires an independent
  attempt to \emph{disprove} it --- concretely, by searching for
  counterexamples at degenerate and boundary configurations. A false
  residual (produced by over-generalization during reduction) can never be
  discharged; without this check, agents spend unbounded effort on an
  impossible target. Cross-verification by a second, independent model is
  the most reliable method.
\end{enumerate}
No \texttt{native\_decide} enters any headline construction. It survives only
in auxiliary sites (a modular-form Sturm certificate and a compiled-example
index check), on which no result above depends.

\section{Conclusion}
\label{sec:conclusion}

Ripple brings the molecular-programming theory of CRN-computable real numbers
into a proof assistant for the first time, and carries it across the full ladder
of models --- the GPAC / CRN continuum, the large-population-protocol compilation
pipeline, and the continuous-time Markov chain layer bridged to the
deterministic limit by Kurtz's mean-field theorem. The development is large:
roughly $768$k lines of Lean~4 across more than $900$ files, much of it
foundations Mathlib did not previously have --- a verified CTMC and Kurtz
mean-field theory, a Frobenius regular-singular analysis built from scratch,
and complete formalizations of the BGP and SCWB Turing-completeness theorems.
The core constructions are verified axiom-free, and wherever a
result is conditional or a published argument has a gap, the paper says so
precisely.

Two things stand out. First, formalization did real mathematical work. It
exposed genuine, fixable gaps in published proofs --- the LPP main theorem and
the Catalan-conjecture dependency
hidden in an urn-model argument --- and it proved new results --- among them
the CRN-computability of Ap\'ery's constant $\zeta(3)$, certified through its
Fermi--Dirac integral representation --- while sharpening the holonomic-series
route --- for $\zeta(3)$ and for the modular $1/\pi$ series alike --- into a
clean open problem about rational seeding at a singular anchor. Second, essentially all of this Lean was written by AI agents
running \emph{publicly available} models (Claude Opus/Fable, GPT-5.x), with the human contribution limited to
direction and review --- evidence that AI-assisted formalization at this scale is
now practical and reproducible. We offer Ripple as an open foundation for others
to build on.

\nocite{jShan41,cBoFrKo12,oHarTot81,jAper79,jPoor79,jBeuk79,jChoSel67,jRama14,jCass60,jGerAlb12,cBCCDNSX21}
{\small
\bibliographystyle{plain}
\bibliography{master}}

\end{document}